# A Statistical Analysis of Sunspot Active Longitudes

A. Malik[1] and M. Bohm[2]
*Department of Space Science, Box 812, 981 28 Kiruna, Sweden*
[1](e-mail: abdmal-6@student.ltu.se)
[2](e-mail: martin.bohm@ltu.se)

**Abstract.** This research work is based on the study of the longitudinal distribution of the most active sunspot zones on the photosphere. Sunspot data has been analyzed for 12 solar cycles (cycles 12-23) separately for northern and southern hemisphere, and the entire solar sphere. The time-longitude diagrams and their corresponding histograms have been plotted to probe the formation, location, longitudinal spread, and lifetime of the most active sunspot longitudes. By the analysis and comparison of the time-longitude diagrams and histograms six active longitudes (≳ 0°, ~90°, ~135°, ~180°, ~270° and ≲ 360°) have been identified out of which three (~90°, ~180° and ~270°) are observed to be most frequent for the whole dataset analyzed. The comparison of the northern and southern hemisphere revealed that the hemispheres do not exhibit very similar kind of behavior. The lifetime and longitudinal spread of sunspot active longitudes is found to be 3-5 Carrington rotations and 20-30° Carrington longitude respectively. This research work also includes the study of the movement of most active sunspot longitudes from higher to lower latitudes in northern hemispheres for six solar cycles (cycles 18-23). For this purpose the formation of sunspot active longitudes has been investigated in four latitudinal belts (40-30°, 30-20°, 20-10° and 10-0°)). It is found that the sunspot active longitudes follow a certain longitudinal pattern during the evolution of the 11-year solar cycle. In the beginning of a solar cycle they seem to appear mostly around two longitudes ~0° and ~270° in the latitudinal belt 40-30°. As the solar cycle proceeds they tend to be stable around two longitudes ~90° and ~270° which are antipodal.

## 1. Introduction

Sunspots emerge in narrow latitudinal belts and move towards the equator as the solar cycle progresses, this pattern is known as the Maunder butterfly diagram. The latitudinal distribution of sunspot activity during the solar cycle is well established but the longitudinal behavior of sunspot activity does not show such a clear pattern (Berdyugina and Usoskin, 2003). An unexplained phenomenon in the magnetic activity of the Sun is its longitudinal non-uniformity. This phenomenon is called active longitudes (Kitchatinov and Olemskoi, 2005). Active longitudes are the longitudes where magnetic activity is enhanced or reoccurs over long durations (Brandenburg and Subramanian, 2005; Ivanov, 2007). Such zones are mostly obvious during the waning phase of the solar cycle. The history of active longitudes is quite old. Carrington was the first man who noticed that sunspots are not randomly distributed over the solar surface but appear in certain heliographic longitudes (Usoskin et al., 2007).

Long-lasting active zones were detected at different Carrington longitudes in the northern and southern hemispheres, and found to rotate with a period different from that of the Carrington system (Usoskin et al., 2007). Many earlier researchers believe that the structure of active longitudes is rigid in Carrington system and their rotation rate is constant. However, some authors say that the rotation rate of active longitudes may change in time (Berdyugina and Usoskin, 2003). The sunspot active longitudes migrate in Carrington longitude and seem to rotate differentially (Usoskin et al., 2007). The migration of active longitudes may be due to changes of the mean latitude of sunspot formation and the differential rotation (Berdyugina and Usoskin, 2003; Berdyugina, 2007).

This work is an effort to understand the longitudinal distribution of sunspot activity for the period 1878-2008 (solar cycles 12 to 23). A statistical analysis of sunspot data for these solar cycles has been made. The formation, location, longitudinal spread, and lifetime of the most active sunspot longitudes have been investigated separately in the northern and southern hemispheres, and also in the entire solar sphere. A comparison has been made to see how the formation of sunspot active longitudes varies in the two hemispheres and in the entire solar sphere. The question whether the sunspot active longitudes follow a certain longitudinal pattern or they have random longitudes in different latitudinal belts during the evolution of the 11-year solar cycle, has also been addressed. This migration pattern of active longitudes from higher to lower latitudes has been analysed for six solar cycles (cycles 18-23).

Active longitudes for representative solar cycles (cycles 12-23) have been identified and an effort has been made to see if there exists any similarity for the active longitudes of one solar cycle to the subsequent solar cycles.

## 2. Data and Processing

### 2.1. Description of the Dataset Analyzed

Two datasets, sunspot dataset and Carrington rotation calendar, were used to carry out this study. The sunspot dataset was downloaded from the NASA Solar Physics web site[1] and Carrington rotation calendar was downloaded from the Association of Lunar & Planetary Observers (ALPO) web site[2].

During the processing of the data we found some irregularities in the data. For instance, no sunspot observations were found for Carrington rotation numbers 339-340, 343, 483, 631, 635-636, 640, 649, 780, 797-798, 801, 1072, 1914 and 2062. The heliographic longitudes range from 0° to 360° but in the dataset a few longitudinal values were observed with Carrington longitude higher than 360°. For example, in cycle 13 and Carrington rotation No. 593 the longitudes 360°.1, 360°.2 and 360°.6 were observed. As they were more close to 0° they were replaced in the dataset by 0.1°, 0.2° and 0.6° respectively. Pelt et al. (2006) also found similar kind of outlier while processing the data downloaded from the same NASA *Solar Physics* web site. He subtracted 360° degrees from the observed higher values (assuming circularity). Both the techniques give similar results.

### 2.2. Data Processing Technique

We have processed the data for 12 solar cycles (1878 to 2008) from Carrington rotation No. 337 to 2067. The Carrington rotation numbers were appended, from Carrington rotation calendar, with the sunspot dataset files according to the rotation start and stop times. To see the distribution of sunspot active longitudes in these solar cycles (cycles 12-23) we have summed up the daily areas of sunspot groups over a Carrington rotation at 10° resolution of Carrington longitude (also called rotation-summed sunspot area). By plotting the rotation-summed sunspot area vs. Carrington longitude (also called the time-longitude diagram) we can easily see the formation and locations of most active regions of sunspots (see Figures 1(a) and 1(c)). According to Ivanov (2003) this technique outlines the intensive sunspot active longitudes much better than rotation maximum sunspot areas or relative numbers of sunspot groups. Also according to Berdyugina and Usoskin (2003) summing up sunspot areas over short time scales of one Carrington rotation allows recovering locations of sunspot clusters which are not smeared by differential rotation and assumed to be constant during one Carrington rotation.

[1] http://solarscience.msfc.nasa.gov/greenwch.shtml
[2] ALPO Solar Section: Carrington Rotation Commencement Dates from Years 1853-2016 (Rotation Numbers -10 to 2172): http://alpo-astronomy.org/

To further verify the formation of most active sunspot longitudes and to separate them from less active regions in the time-longitude diagram, histograms have been plotted (see Figures 1(b) and 1(d)). These histograms, which show the area of sunspot active regions (in millionth of visible hemisphere (mvh)) vs. Carrington longitudes, have been plotted with 10° interval of Carrington longitude so we have 36 total bins in each histogram. To produce these histograms all the sunspot areas have been summed up over all the Carrington rotations under consideration at 10° interval of Carrington longitude. This index used to produce histograms was introduced by Ivanov (2007) and is called the total rotation-summed sunspot area. To further clarify, this total rotation-summed sunspot area is not the sum of the daily areas of sunspot groups over a single Carrington rotation instead it is the sum of the daily areas of sunspot groups over all the Carrington rotations in a particular cycle at 10° interval of Carrington longitude. The histograms also show the smooth curves which have been obtained by taking an average of sunspot areas in 5 following longitudinal bins. Ivanov (2003) obtained these smooth curves by averaging sunspot areas over moving intervals equal to 5 Carrington rotations. Both the techniques to plot the smooth curves give similar results with minor differences. The horizontal line in each histogram represents the mean value of the total rotation-summed sunspot area.

These time-longitude diagrams and histograms with smooth curves have been plotted for twelve solar cycles (12-23) separately for northern hemisphere (Section 3.1), southern hemisphere (Section 3.2), and for the entire solar sphere (northern and southern hemisphere together) (Section 3.3). Besides this the time-longitude diagrams and histograms with smooth curves have also been plotted to see the migration pattern of sunspot active longitudes from higher to lower latitudes and it has been analyzed only for northern hemispheres of six solar cycles (18-23) (Section 3.4). This migration pattern of sunspot active longitudes from higher to lower latitudes have been studied from 40° to 0° north latitude in four latitudinal belts (40-30°, 30-20°, 20-10° and 10-0°). Both the time-longitude diagrams and histograms have been plotted separately for each latitudinal belt.

## 3. Analysis

The plots for northern and southern hemisphere for 12 solar cycles (12-23) contain the sunspot data from 40° to 0° north latitude and -40° to 0° south latitude respectively. The locations of sunspot active zones, their longitudinal spread, and life time have been analyzed. Only those regions have been taken as active regions whose smooth curves are either above the mean horizontal line or at least peak of the curve lies at the mean horizontal line in the histogram. So the mean horizontal line acts as threshold for identification of the sunspots active zones. There are so many plots for the northern hemisphere, southern hemisphere, entire solar sphere, and for different latitudinal belts that all cannot be shown here. So we have shown some representative plots and the results from all of the plots have been summarized in the respective tables (Table 1 to 7). We have found that the most active sunspot regions tend to exist near six Carrington longitudes and these are $\gtrsim$ 0°, ~90°, ~135°, ~180°, ~270°, and $\lesssim$ 360°.

**Table 1** Locations of most active sunspot longitudes in Northern (N) hemisphere for cycles 12-23

| Solar Cycle (SC) No. | Observed at ≳ 0° | Observed at ~90° | Observed at ~135° | Observed at ~180° | Observed at ~270° | Observed at ≲ 360° |
|---|---|---|---|---|---|---|
| SC-12N | ~30° | | | ~180° | ~250° | |
| SC-13N | ~45° | | | ~180° | | ~310° |
| SC-14N | | | | ~180° | ~270° | ~360° |
| SC-15N | ~0° | ~110° | | | ~290° | |
| SC-16N | ~30° | | ~135° | ~180° | ~290° | |
| SC-17N | | ~110° | | ~190° | | ~310° |
| SC-18N | | ~90° | | 180 | ~270° | |
| SC-19N | | ~90° | | ~180° (below threshold) | ~290° | |
| SC-20N | | ~90° | | ~180° (below threshold) | ~270° | |
| SC-21N | | ~110° | | ~180° | | ~330° |
| SC-22N | ~20° | | ~135° | | ~270° | ~360° |
| SC-23N | ~20° | ~90° | | ~180° | ~290° | ~360° |

### 3.1. Northern Hemisphere

The time-longitude diagrams and histograms for northern hemispheres for solar cycle 18 and 20 are shown in Figure 1. By comparison of time-longitude diagrams and corresponding histograms for the northern hemisphere we have identified the most active sunspot longitudes for 12 solar cycles (12-23) that are summarized in Table 1.

Table 1 shows the six most active longitudes (observed at ≳ 0°, ~90°, ~135°, ~180°, ~270°, and ≲ 360°) although one of them (at ~135°) is hard to see. The active longitudes around 0° and 360° do not seem to appear consistently. We don’t see any active longitude at ~90° in solar cycles 12-14, 16 and 22. However, in cycles 18-20 and 23 the active longitudes at ~90° can be seen clearly (see Figure 1 for cycles 18 and 20). During analysis we noted that in cycles 15, 17 and 21 the sunspots tend to cluster around 90° and apparently we observe an active zone at this longitude.

The active longitudes appear more frequently at ~180° in 12 solar cycles analyzed for northern hemisphere. Only cycles 15 and 22 do not show any active zone at or close to this longitude. Solar cycles 19-20 show active longitudes with curves below the threshold value (see Figure 1(d)) but in most cases we may say that the active zones have higher tendency to appear at this longitude than at others. Cycles 12-14, 16, 18, 21 and 23 are good examples to confirm the existence of sunspots active zones at ~180°. In cycle 17 the peak of the curve was slightly away from 180° but the analysis of time-longitude diagrams for this cycle showed a tendency of occurrence of sunspots at this longitude.

Sunspot active longitudes in solar cycles 12, 15-16, 19 and 23 were not well centered at 270° but we observed a peak curve close to this longitude. Cycles 13, 17 and 21 do not show any active zone at this longitude. However, cycles 14, 18, 20, and 22 clearly show active regions at this longitude (see Figure 1 for cycles 18 and 20).

Although in the northern hemisphere the most frequently observed active longitudes are at ~90°, ~180°, and ~270° but they do not seem to appear very consistently throughout all the solar cycles analyzed. The sunspot active longitudes show some variations in their positions from cycle to cycle.

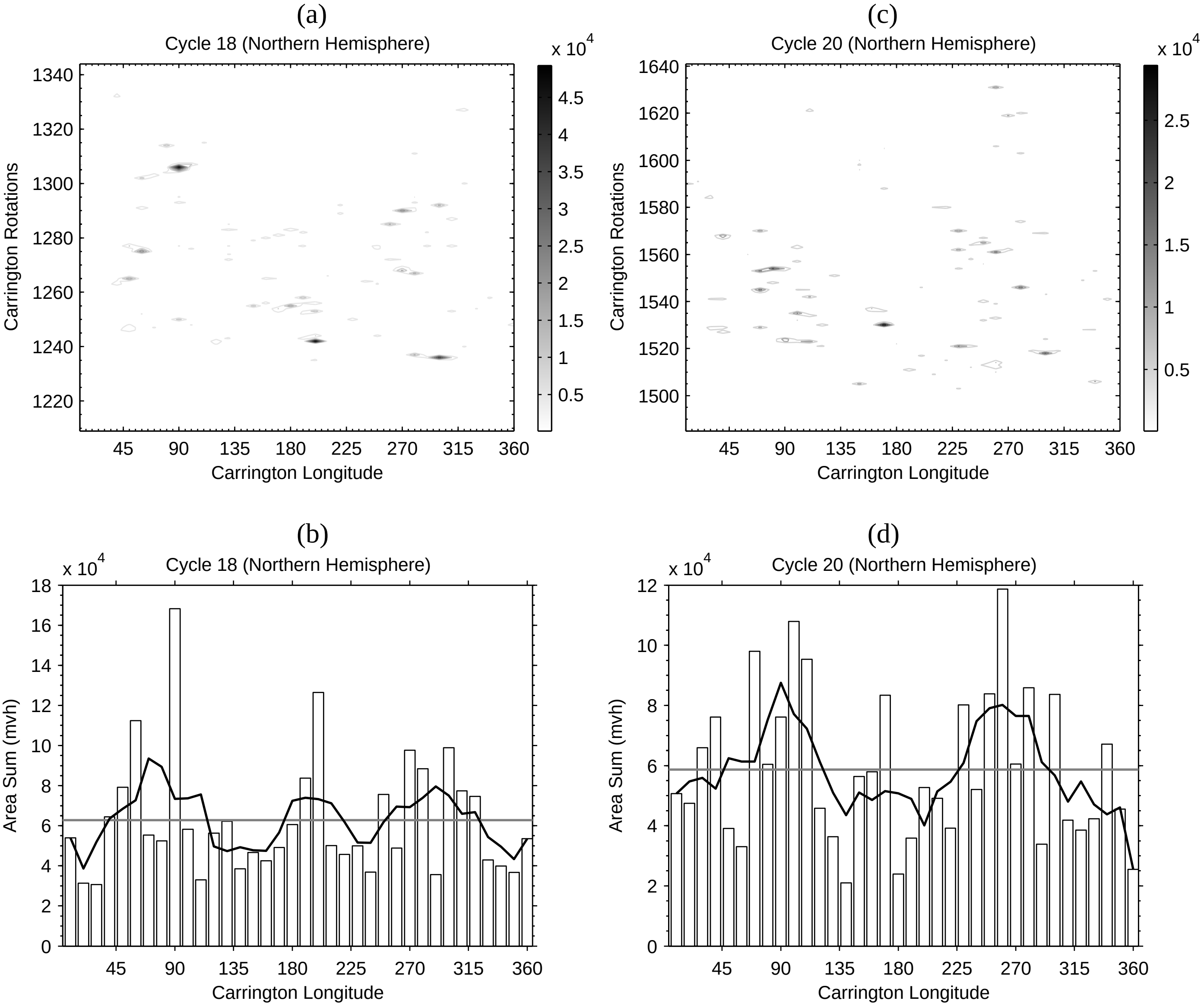

**Figure 1** Time-longitude diagrams and histograms for cycle 18 (a, b) and for cycle 20 (c, d) in the northern hemisphere. The color bar shows the rotation-summed sunspot area in millionth of visible hemisphere (mvh).

**Table 2** Locations of most active sunspot longitudes in Southern (S) hemisphere for cycles 12-23

| Solar Cycle (SC) No. | Observed at ≳ 0° | Observed at ~90° | Observed at ~135° | Observed at ~180° | Observed at ~270° | Observed at ≲ 360° |
|---|---|---|---|---|---|---|
| SC-12S | | ~70° | | | ~240° | |
| SC-13S | | ~70° | | | ~270° | |
| SC-14S | | ~70° | | ~180° | | ~330° |
| SC-15S | ~0° | | ~135° | ~210° | | |
| SC-16S | | ~90° | | | ~270° | |
| SC-17S | ~40° | | | ~190° | ~270° | |
| SC-18S | ~0° | ~90° | | | | |
| SC-19S | ~0° | ~90° | | ~180° | | ~330° |
| SC-20S | | ~90° | | ~180° | | ~330° |
| SC-21S | | ~90° | | ~180° | ~270° | ~360° |
| SC-22S | ~0° | ~90° | | ~180° | | ~330° |
| SC-23S | | | ~135° | | ~270° | |

### 3.2. Southern Hemisphere

The time-longitude diagrams and histograms for southern hemispheres are shown in Figure 2 for solar cycle 18 and 19. By comparison of time-longitude diagrams and corresponding histograms for southern hemisphere we have identified the most active sunspot longitudes for 12 solar cycles (12-23) that are summarized in Table 2.

In southern hemisphere the active longitudes at ≳ 0°, ~135°, and ≲ 360° do not seem to appear more frequently than at ~90°, ~180°, and ~270°. We do not see any active longitudes at or near 90° in solar cycles 15, 17 and 23. Cycles 18-22 show sunspots active zones at this longitude with quite consistency. For cycles 12-14 the time-longitude diagrams revealed that sunspots tend to cluster close to 90° so naturally we observe a peak near this longitude. We note that the most frequently observed active longitude in southern hemisphere is at ~90°.

The 2$^{nd}$ most frequently observed active longitude in southern hemisphere is at ~180. We noticed that the sunspots active zones do not appear at 180° in cycles 12-13, 16, 18 and 23 whereas they were well seen in cycles 14, and 19-22. Cycles 15 and 17 showed peaks slightly away from 180°.

The third frequently observed active longitude in southern hemisphere is at ~270° and it can be seen in cycles 12-13, 16-17, 21 and 23.

The analysis of time-longitude diagrams for northern and southern hemisphere shows that the lifetime of sunspot active longitudes seems to vary between 3 and 5 Carrington rotations. The longitudinal spread of active longitudes lies between 20-30° and it is shown by two vertical lines in southern hemisphere of solar cycle 18 (Figure 2(a)). It seems that both the hemispheres have the same lifetime and longitudinal spread of active longitudes.

### 3.3. Entire Solar Sphere

The data for entire solar sphere have been processed as a whole to see that how differently each hemisphere (northern and southern) behaves from the entire solar sphere. Here we have not shown the time-longitude diagrams and histograms for the entire sphere and only the results have been summarized in Table 3.

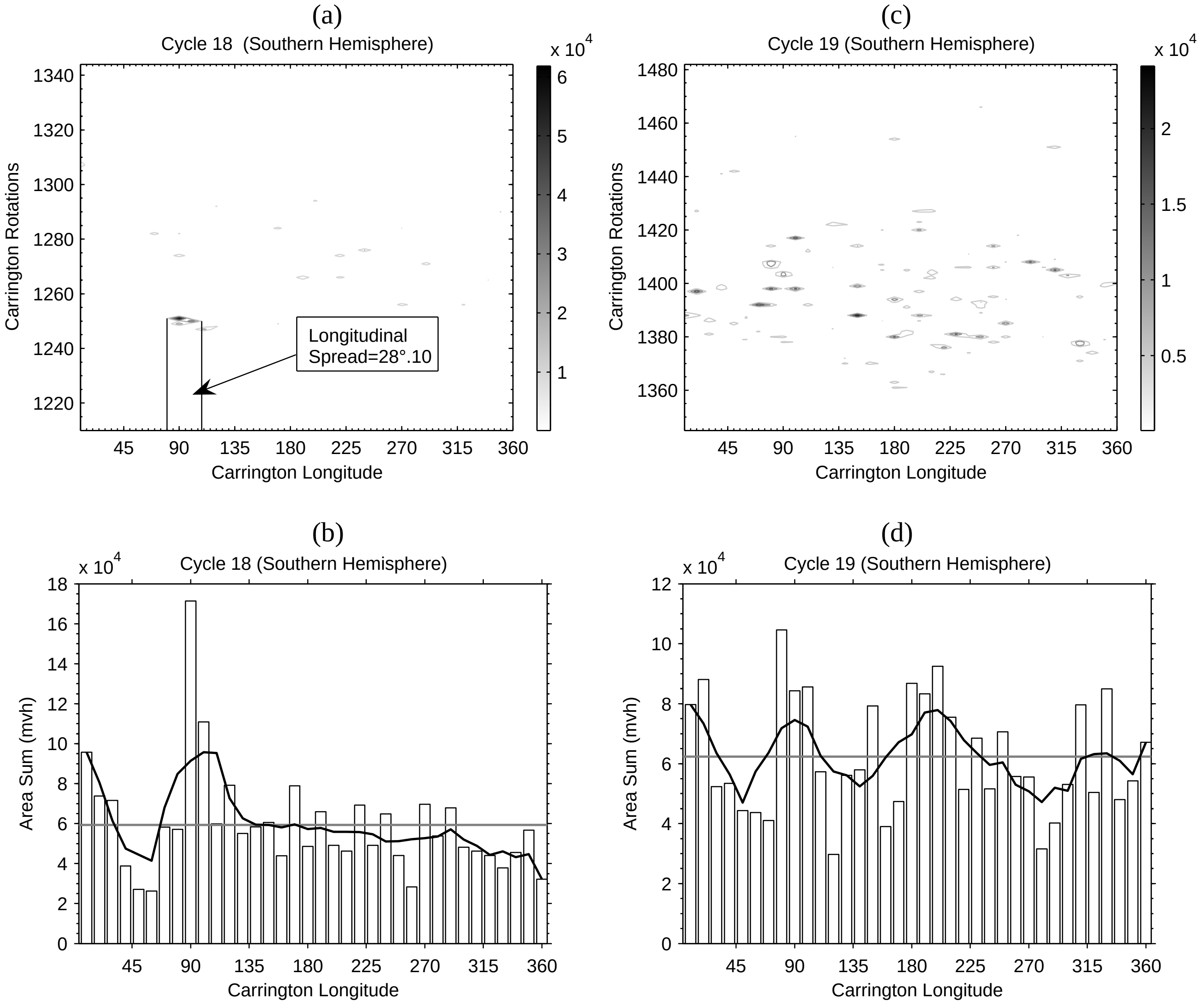

**Figure 2** Time-longitude diagrams and histograms for cycle 18 (a, b) and for cycle 19 (c, d) in the southern hemisphere. The color bar shows the rotation-summed sunspot area in millionth of visible hemisphere (mvh). The vertical lines in solar cycle 18 (a) show the longitudinal spread of a sunspot active longitude.

**Table 3** Locations of most active sunspot longitudes in entire solar sphere for cycles 12-23

| Solar Cycle (SC) No. | Observed at ≳ 0° | Observed at ~90° | Observed at ~135° | Observed at ~180° | Observed at ~270° | Observed at ≲ 360° |
|---|---|---|---|---|---|---|
| SC-12 | ~45° | | | | ~250° | |
| SC-13 | ~45° | | | | | ~320° |
| SC-14 | ~45° | | | ~180° | | ~330° |
| SC-15 | ~0° | ~110° | | | | |
| SC-16 | ~45 | ~110° | | ~210° | ~270° | |
| SC-17 | | ~110° | | ~180° | ~270° | |
| SC-18 | ~0° | 90° | | ~180° | ~270° | |
| SC-19 | ~0° | 90° | | ~190° | | ~320° |
| SC-20 | | 90° | | ~180° (below threshold) | ~270° | |
| SC-21 | | ~110° | | ~180° | ~270° | |
| SC-22 | ~0° | ~90° | | ~180° | ~270° | |
| SC-23 | | | | | ~270° | |

An active longitude seems to exist at ~0° in some solar cycles (15, 18-19, and 22) and in others (12-14, and 16) we observed that the sunspots tend to cluster near 0°. The active longitude observed at ~0° does not seem to appear constantly but we do find a tendency of its appearance at this longitude. The tendency of appearance of an active longitude close to 360° is very little.

Cycles 12-14 and 23 do not show any active longitude at ~90° but they can be seen clearly in cycles 18-20 and 22. Cycles 15-17 and 21 show active longitudes with peak slightly away from 90°.

The active longitudes at ~180° (in cycles 14, 16-22) and 270° (in cycles 12, 16-18, and 20-23) can also be seen frequently.

### 3.4. Higher to Lower Latitude Movement of Active Longitudes

To see the formation of active longitudes in specific latitudinal belts we have divided the northern hemisphere into four belts (40-30°, 30-20°, 20-10° and 10-0°) and the length of each latitudinal belt is 10°. According to Ivanov (2007) sunspot formation zones appear in a broad band of heliolatitudes up to 10-20°.The time-longitude diagrams and histograms of these latitudinal belts are shown for cycle 18 in Figures 3 and 4.The results obtained by studying the time-longitude diagrams and corresponding histograms for four latitudinal belts in northern hemisphere (for cycles 18-23) have been summarized in Tables 4-7.

We observe two active longitudes in latitudinal belt 40-30° mostly seem to appear at ~270° and near 0° (see Table 4 and Figures 3(a), 3(b)). The active longitude at ~270° appears consistently at cycles 18-22 whereas the active longitude at ≳ 0° does not seem to appear with constancy at cycles 18-23. In this belt the active longitudes identified at or near ~90°, and ~180° are not so frequent.

Latitudinal belt 30-20° (see Table 5) shows a different behavior than the latitudinal belt 40-30°. Here we detect three more frequent active longitudes at ~90°, ~180° and ~270°. We also get the impression that in this latitudinal belt the sunspots tend to cluster around these three longitudes (Figures 3(c) and 3(d)). In this belt the active longitude at ~270° seems more persistent and frequent than at ~90° and ~180°.

We have found that the mean latitude of sunspot formation for solar cycles 18-23 is approximately 16°. Table 6 shows that in the latitudinal belt 20-10° the two active longitudes at ~90° and particularly at ~270° seem to appear constantly and are 180° degree apart. A similar kind of behavior can also be seen in latitudinal belt 10-0° (see Table 7 and Figures 4(c), 4(d)). Figures 4(a) and 4(b) for cycle 18 and latitudinal belt 20-10° shows three active longitudes (at ~90°, ~180° and ~270°) while in other cycles for this belt we mostly observe only two active longitudes (at ~90° and ~270°).

**Table 4** Locations of most active sunspot longitudes in latitudinal belt 40-30° north for cycles 18-23

| Solar Cycle (SC) No. | Observed at ≳ 0° | Observed at ~90° | Observed at ~180° | Observed at ~270° | Observed at ≲ 360° |
|---|---|---|---|---|---|
| SC-18 | ~30° | | | ~250 | ~330° |
| SC-19 | ~0° | | | ~270° | |
| SC-20 | | | ~180° | ~270° | ~360° |
| SC-21 | ~0° | | | ~250° | |
| SC-22 | | | | ~270° | |
| SC-23 | ~45° | ~90° | | | |

**Table 5** Locations of most active sunspot longitudes in latitudinal belt 30-20° north for cycles 18-23

| Solar Cycle (SC) No. | Observed at ≳ 0° | Observed at ~90° | Observed at ~180° | Observed at ~270° | Observed at ≲ 360° |
|---|---|---|---|---|---|
| SC-18 | ~45° | | ~180° | ~290° | |
| SC-19 | | | ~180° | | ~360° |
| SC-20 | | ~90° | | ~270° | |
| SC-21 | | ~90° | ~180° | ~270° | |
| SC-22 | | ~90° | | ~270° | |
| SC-23 | | ~90° | ~180° | ~270° | |

**Table 6** Locations of most active sunspot longitudes in latitudinal belt 20-10° north for cycles 18-23

| Solar Cycle (SC) No. | Observed at ≳ 0° | Observed at ~90° | Observed at ~180° | Observed at ~270° | Observed at ≲ 360° |
|---|---|---|---|---|---|
| SC-18 | | ~90° | ~180° | ~270° | |
| SC-19 | | ~90° | | ~270° | |
| SC-20 | | ~90° | | ~270° | |
| SC-21 | | ~90° | | ~270° | |
| SC-22 | ~0° | ~90° | | ~255° | ~360° |
| SC-23 | | | ~180° | ~270° | |

**Table 7** Locations of most active sunspot longitudes in latitudinal belt 10-0° north for cycles 18-23

| Solar Cycle (SC) No. | Observed at ≳ 0° | Observed at ~90° | Observed at ~180° | Observed at ~270° | Observed at ≲ 360° |
|---|---|---|---|---|---|
| SC-18 | | ~90° | | ~270° | |
| SC-19 | | ~90° | | ~270° | |
| SC-20 | | ~90° | | ~270° | |
| SC-21 | | | ~180 | | |
| SC-22 | ~0° | ~90° | | | ~360° |
| SC-23 | | ~90° | | ~270° | |

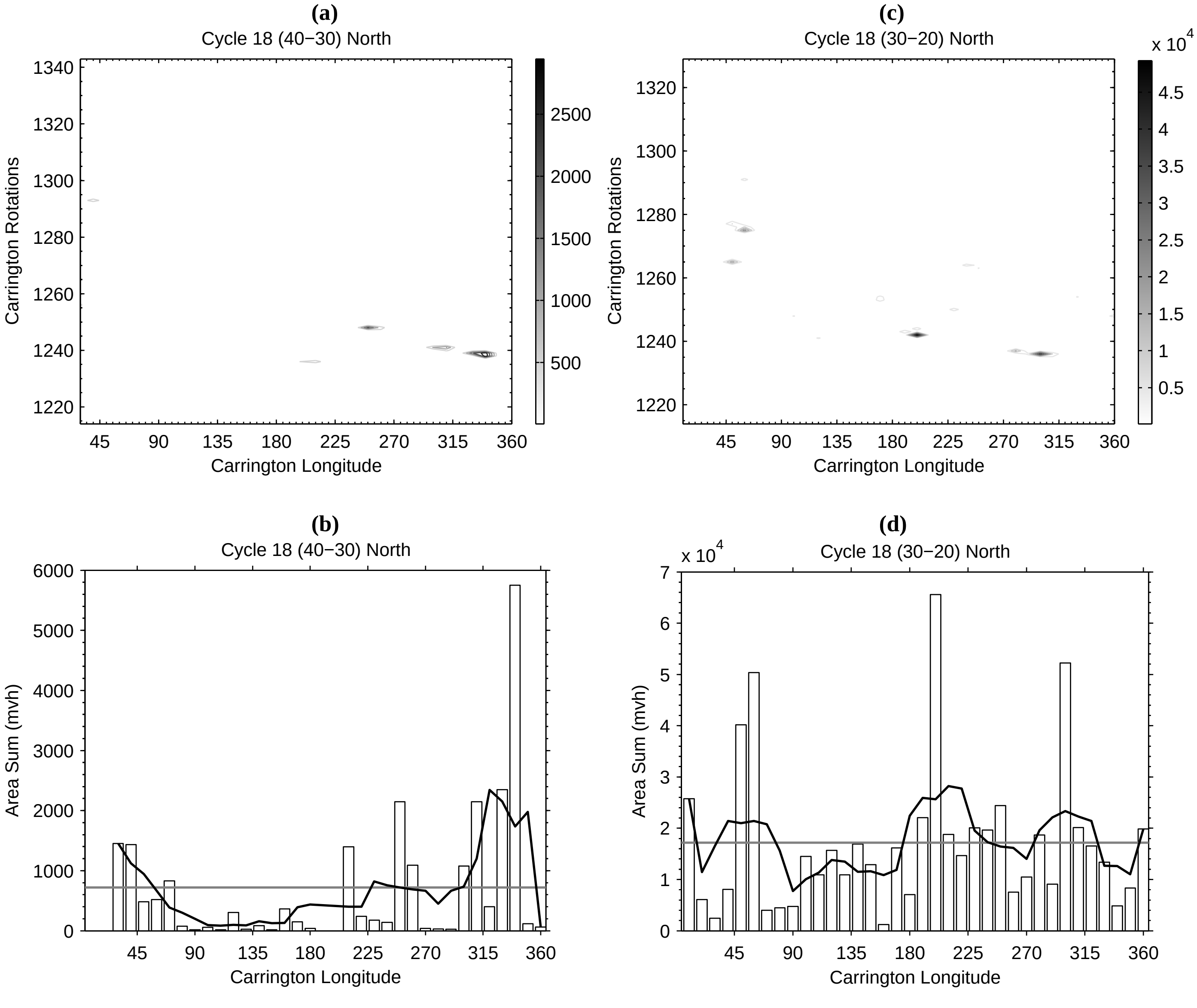

**Figure 3** Time-longitude diagrams and histograms of solar cycle 18 for latitudinal belt 40-30° (a, b) and for latitudinal belt 30-20° (c, d) in the northern hemisphere. The color bar shows the rotation-summed sunspot area in millionth of visible hemisphere (mvh).

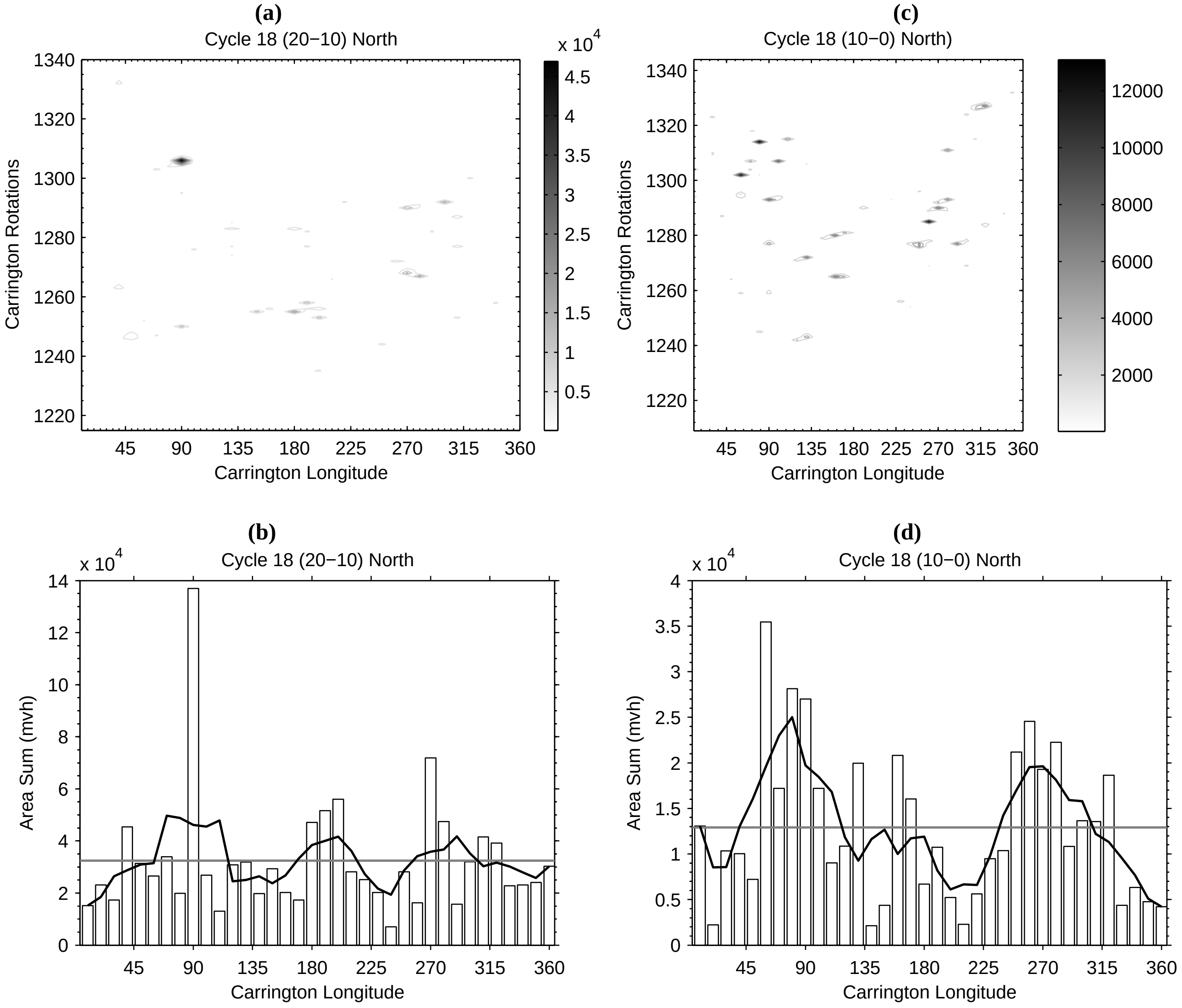

**Figure 4** Time-longitude diagrams and histograms of solar cycle 18 for latitudinal belt 20-10° (a, b) and for latitudinal belt 10-0° (c, d) in the northern hemisphere. The color bar shows the rotation-summed sunspot area in millionth of visible hemisphere (mvh).

## 4. Discussion and Conclusion

The results found regarding the locations and spread of active longitudes fairly agree with the studies of Ivanov (2003 and 2007) for analysis of northern hemisphere, southern hemisphere, and for the entire solar sphere. Only the lifetime of sunspot active longitudes found in our analysis (~3-5 Carrington rotations) does not agree with the lifetime of active longitudes found by Ivanov (2007) and according to his studies it is 15-20 Carrington rotations. Also according to various studies (e.g., Berdyugina and Usoskin 2003; Ivanov 2003; Usoskin, Berdyugina, and Poutanen, 2005; Berdyugina 2007; Usoskin et al., 2007) the active longitudes appear antipodly in pairs that can be well observed in some of the time- longitude diagrams and corresponding histograms for northern hemisphere, southern hemisphere, entire solar sphere, and for the latitudinal belts 20-10° & 10-0°.

We note that the northern and southern hemispheres do not show very similar kind of behavior when we compare the active longitudes in these hemispheres. The comparison shows that the sunspot active longitudes at ~180° and ~270° tend to appear more constantly in northern hemisphere than in southern hemisphere (Table 1 and 2). The sunspot active longitude at ~90° appears more frequently in southern hemisphere than in northern hemisphere. The number of active longitudes and their locations are not always the same in both the hemispheres for a particular cycle. For example in SC-18N we observe three active longitudes (at ~90°, ~180°, and ~270°) while in SC-18S we observe only two active longitudes (at ~0°, and ~90°). Thus we can say that an active longitude appearing at a particular longitude in the northern hemisphere may not necessarily appear at the same longitude in the southern hemisphere.

By the analysis of the northern hemisphere, southern hemisphere, and the entire solar sphere we find that three active longitudes at ~90°, ~180° and ~270° appear most frequently from cycles 12-23. However in cycles 12-14 the active longitude at ~90° does not seem to appear in northern hemisphere and entire solar sphere. The northern and southern hemisphere also differs from the entire solar sphere in terms of the number of active longitudes. For example in SC-15N the number of active longitudes are three (at ~0°, ~110, ~290) while in the entire SC-15 the number of active longitudes are two (at ~0°, ~110°). Also the positions of active longitudes in the northern and southern hemisphere are not always same as in the entire solar sphere.

From the analysis of six solar cycles (18-23) we reach the conclusion that in the beginning of a solar cycle the active longitudes appear more frequently at ~0° and ~270° than at other longitudes. As the solar cycle proceeds the active longitudes are more likely to appear at ~90°, ~180° and ~270° in latitudinal belt 30-20°. From latitude 20-0° the active longitudes mostly appear at ~90° and ~270° which are antipodal.

The longitudinal spread of active longitudes observed during our analysis seems to exist between 20-30° Carrington longitude. According to Ivanov (2007) this spread appears to be mainly the result of ignoring the sunspot proper motions and on the photosphere, sunspots rotate differentially and, during their lifetime (~2-3 rotations), they can move in heliolongitude by 15-20° away from their origin.

In summary we may conclude as follows:

1. We have observed six active longitudes around ≳ 0°, ~90°, ~135°, ~180°, ~270° and ≲ 360° and out of six only three active longitudes around ~90°, ~180° and ~270° are more constant and frequent throughout all the solar cycles analyzed for northern hemisphere, southern hemisphere and entire solar sphere.

2. Northern and southern hemisphere does not behave similarly and their behavior also differs from the entire solar sphere.

3. The longitudinal spread of active longitudes seems to vary between 20-30° Carrington longitude.

4. The lifetime of active longitudes seems to vary between 3-5 Carrington rotations which is slightly larger than the lifetime of the long-lasting individual sunspots.

5. Active longitudes seem to change their positions in different latitudinal belts during the evolution of a solar cycle. As in the beginning of a solar cycle the active longitudes mostly seem to appear around ~0° and ~270°. With the passage of time they start appearing around ~90°, ~180° and ~270° and ultimately they become more stable around two active longitudes ~90° and ~270° which are antipodal.